\documentclass[11pt]{article}

\usepackage{amsmath,amssymb}

\usepackage{epsfig}
\usepackage{graphicx}               
\usepackage{url}
\usepackage{hyperref}

\newcommand{\nc}{\newcommand}

\nc{\beq}{\begin{equation}}
\nc{\eeq}{\end{equation}}
\nc{\beqa}{\begin{eqnarray}}
\nc{\eeqa}{\end{eqnarray}}
\nc{\bea}{\begin{eqnarray}}
\nc{\eea}{\end{eqnarray}}
\nc{\ra}{\rightarrow}
\nc{\lsim}{\begin{array}{c}\,\sim\vspace{-21pt}\\< \end{array}}
\nc{\gsim}{\begin{array}{c}\sim\vspace{-21pt}\\> \end{array}}
\nc{\Tr}{{\rm Tr}}
\nc{\slsh}{\slash\hspace*{-0.22cm}}

\def\be{\begin{equation}}
\def\ee{\end{equation}}
\def\bea{\begin{eqnarray}}
\def\eea{\end{eqnarray}}
\def\bit{\begin{itemize}}
\def\eit{\end{itemize}}

\title{
\vspace*{-2.3cm}
\begin{flushright}
\normalsize{
SLAC-PUB-14956
  }
\end{flushright}
\vspace{1.5cm}
\Large
\textbf{
Reach the Bottom Line of the Sbottom Search
}\vspace*{1.0cm}
}

\author{Ezequiel \'Alvarez$^{a,b}$ and Yang Bai$^{b}$
\vspace{5mm}
\\
$^{a}$\normalsize\emph{CONICET, INFAP and Departamento de F\'{\i}sica, FCFMN, Universidad Nacional de San Luis} \\
\normalsize\emph{Av. Ej\'ercito de los Andes 950, 5700, San Luis, Argentina} \\
\vspace{-2mm} \\
$^{b}$\normalsize\emph{SLAC National Accelerator Laboratory, 2575 Sand Hill Road, Menlo Park, CA 94025, USA} 
}

\date{}

\begin{document}
\setcounter{page}{0}
\maketitle

\vspace*{1cm}
\begin{abstract}
We propose a new search strategy for directly-produced sbottoms at the LHC with a small mass splitting between the sbottom and its decayed stable neutralino. Our search strategy is based on boosting sbottoms through an energetic initial state radiation jet. In the final state, we require a large missing transverse energy and one or two $b$-jets besides the initial state radiation jet. We also define a few kinematic variables to further increase the discovery reach. For the case that the sbottom mainly decays into the bottom quark and the stable neutralino, we have found that even for a mass splitting as small as 10 GeV sbottoms with masses up to around 400 GeV can be excluded at the 95\% confidence level with 20 inverse femtobarn data at the 8 TeV LHC.
\end{abstract}

\thispagestyle{empty}
\newpage

\setcounter{page}{1}

\baselineskip18pt

\vspace{-3cm}
\section{Introduction}
\label{sec:introduction}
The Large Hadron Collider (LHC) has already entered an exciting era for understanding the physics related to the electroweak symmetry breaking. The existence of the Standard Model (SM) Higgs boson will be confirmed or disconfirmed this year, 2012. If a light Higgs boson is indeed discovered this year, the next big question is to understand the physics beyond the standard model, which makes the models with a light Higgs boson natural and (or) explains the large hierarchy between the fundamental Planck scale and the electroweak scale.  One of the beyond-the-standard models that naturally explains the lightness of the Higgs boson is supersymmetry (SUSY). To cancel the large radiative corrections to the Higgs mass in the SM from the top quark loop without fine tuning, the top superpartners (stops) need to be light enough~\cite{Dimopoulos:1995mi, Cohen:1996vb}. There are increasing amount of interests in the literature on generating SUSY spectra with a light stop from model-building or improving stop searches at the LHC from collider studies (see Ref.~\cite{Lodone:2012kp} for a recent review). Noticing that the left-handed stop and the left-handed sbottom belong to a weak doublet, their masses should be naturally comparable. Light stops below 1 TeV implicitly imply at least one sbottom below around 1 TeV. Therefore, searching for sbottoms is equally important for us to understand the electroweak symmetry breaking in SUSY~\cite{Lee:2012sy}. 

Sbottoms could be directly produced in pairs at hadron colliders from their QCD interaction. With an unbroken R-parity and a stable neutralino $\tilde{\chi}$ as the lightest super-symmetric particle (LSP), the dominant decay channel for the lighter sbottom is $\tilde{b}_1 \rightarrow b\tilde{\chi}$. For the parameter space with a large mass difference between sbottom and neutralino, the standard signature for the direct sbottom search is two jets containing at least one $b$-jet plus a large missing energy $E_T^{\rm miss}$. By requiring at least one $b$-jet in the final state, the existing search from CDF with 2.65 fb$^{-1}$ of integrated luminosity has excluded sbottom masses up to 230 GeV for neutralino masses below 70 GeV at 95\% confidence level (C.L.)~\cite{Aaltonen:2010dy}. With a larger luminosity of 5.2 fb$^{-1}$ and requiring two $b$-jets, the D0 collaboration has set a 95\% C.L. lower limit on the sbottom mass to be 247 GeV for the neutralino mass below 40 GeV~\cite{Abazov:2010wq}. At the 7 TeV LHC, the ATLAS collaboration has performed a search of the sbottom particle in 2.05 fb$^{-1}$ and extended the limits on the sbottom mass to be 390 GeV for neutralino masses below 60 GeV~\cite{Aad:2011cw}.

The existing searches on sbottoms (i.e., Ref.~\cite{Aad:2011cw}) do not cover a wide range of sbottom and neutralino mass parameter space when their masses are close to each other. In this paper, we explore new search strategies to fill this gap and hope to cover the region for the mass difference of sbottom and neutralino reaching to almost the bottom quark mass. For this compressed or squeezed spectrum, the traditional search strategy by requiring a large transverse missing energy with two $b$-jets (see Ref.~\cite{Alwall:2011zm} and \cite{AdeelAjaib:2011ec} for recent collider studies) is not optimized because $E_T^{\rm miss}$ and the $p_T$ of the leading $b$-jet decrease as the mass splitting decreases. However, if one requires one additional hard jet from the initial state radiation (ISR), the situation will be different because the two neutralino's in the decaying products are not only boosted but also move in a direction close to each other. As a result, the total transverse missing energy can be sufficiently large to reduce the SM backgrounds.

One might think that the mono-jet search results can be used to cover the squeezed spectrum~\cite{Aaltonen:2012jb, CMSmonojet}. This is indeed the case for a light sbottom with a large production cross section. However, the mono-jet search is still not the optimized one to cover a large fraction of the sbottom-neutralino mass parameter space because of the large $Z$ plus jets background. To extend the search limit of sbottom, we propose a new search strategy in this paper by requiring one energetic non $b$-tagged jet from the ISR, a large transverse missing energy and one (or two) $b$-tagged jet with a modest transverse energy. Other than the proposal of this general search strategy, we also explore additional good kinematic variables to further reduce the SM backgrounds. 

Our paper is organized as following. We first study the boost of sbottoms from an ISR jet and discuss the general search strategy in Section~\ref{sec:boost}. Then, we explore and present three additional variables to further cut the SM backgrounds in Section~\ref{sec:strategy}. In Section~\ref{sec:reach}, we show the discovery reach and compare our search strategy with the existing strategy for the squeezed spectrum. We discuss other issues and conclude our paper in Section~\ref{sec:conclusion}.

\section{Boosting the Sbottom from an ISR jet}
\label{sec:boost}
One of the existing searches of sbottoms at ATLAS is based on the direct production of a pair of sbottoms using the decay channel $\tilde{b}_1 \rightarrow b + \tilde{\chi}$~\cite{Aad:2011cw}. In their search, they require two $b$-tagged jets with $p_T(b_1) > 130$~GeV ($b_1$ representing the leading $b$-jet) and $p_T(b_2) > 50$~GeV on top of the missing energy cut $E_T^{\rm{miss}} > 130$~GeV.  One variable called the boost-corrected contransverse mass, $m_{\rm CT}\equiv ([E_T(v_1) + E_T(v_2)]^2 - [{\bf p_T}(v_1) - {\bf p_T}(v_2)]^2)^{1/2}$, has been introduced to further reduce the backgrounds. Based on 2.05 fb$^{-1}$ of data at $\sqrt{s}= 7$~TeV, sbottom masses up to 390 GeV are excluded at 95\% C.L. for $m_{\tilde{\chi}} < 60$~GeV and $\mbox{Br}(\tilde{b}_1 \rightarrow b \tilde{\chi}) = 100\%$. From Fig.~2 of Ref.~\cite{Aad:2011cw}, one can see that the current search strategy at ATLAS has not yet covered a wide region of parameter space for $\Delta m \equiv m_{\tilde{b}_1} - m_{\tilde \chi}$ between the bottom quark mass $m_b$ and around 150 GeV.
 
\begin{figure}[h!t]
\begin{center}
\includegraphics[width=0.45\textwidth]{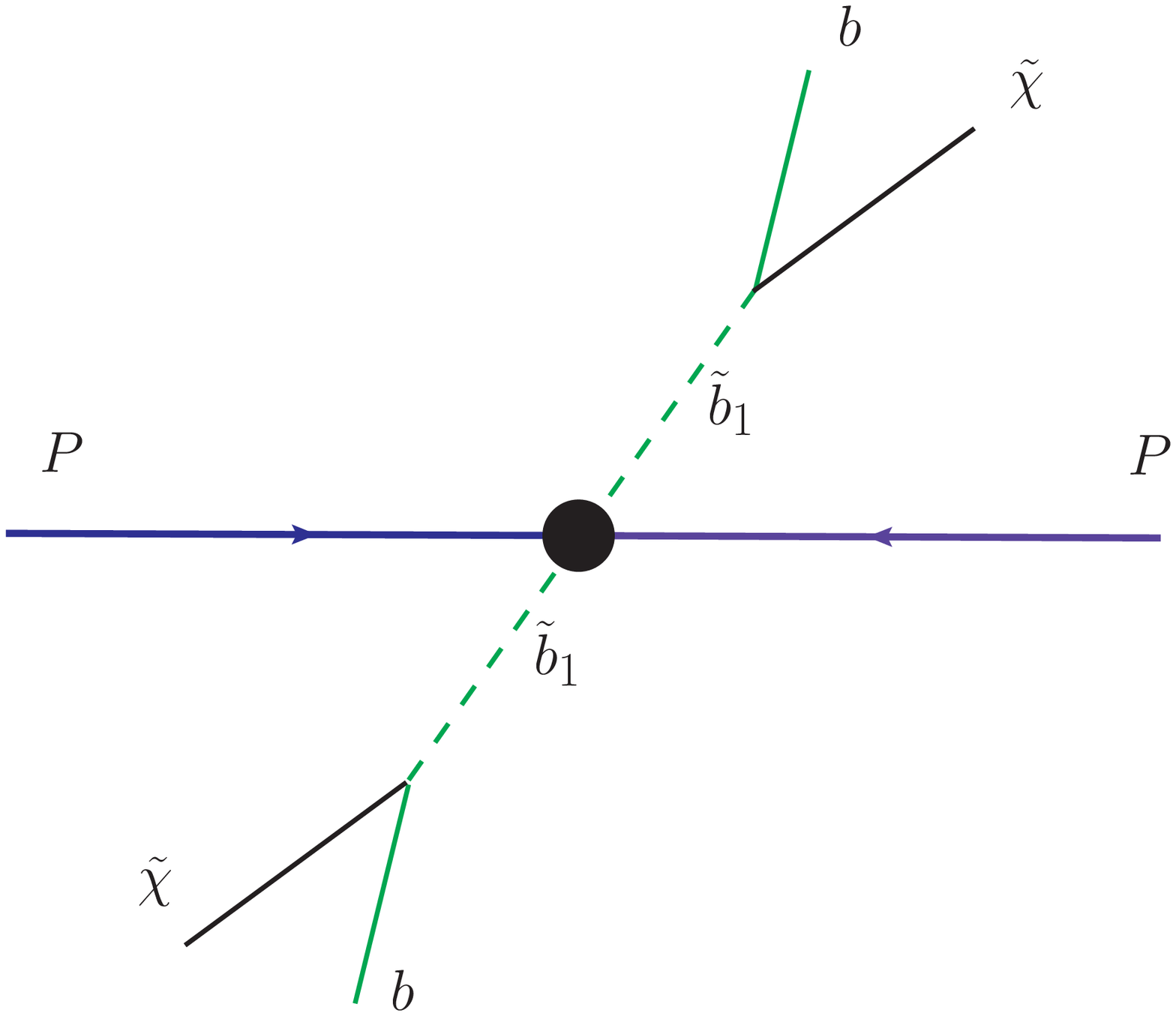}
\hspace{6mm}
\includegraphics[width=0.45\textwidth]{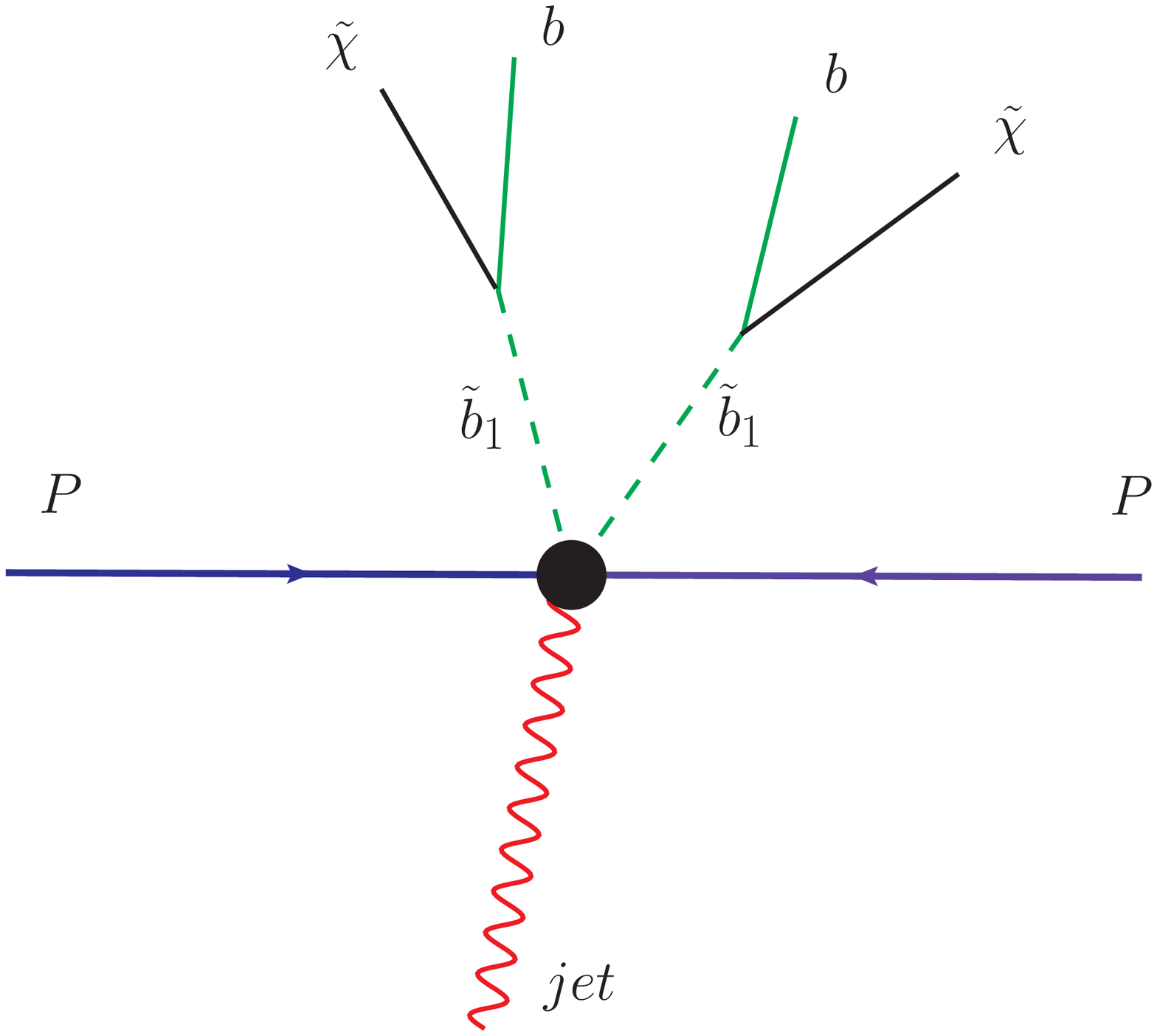}
\caption{Left panel: a schematic plot for the pair-production of sbottoms at the LHC. Right panel: the pair-production of sbottoms together with an energetic ISR jet.}
\label{fig:isr}
\end{center}
\end{figure}

The reason for the limitation of the ATLAS searches on the squeezed spectrum is two-fold. First, when $\Delta m$ is small compared to the neutralino mass, the momentum of $\tilde \chi$ in the rest frame of $\tilde{b}_1$ is $\Delta m$ after neglecting the bottom quark mass $m_b$. If there is no additional jet to boost the two-$\tilde{b}_1$ system in the transverse direction, as illustrated in the left panel of Fig.~\ref{fig:isr},  the missing transverse energy $E_T^{\rm miss}$ should be bounded by $2\Delta m$, with the upper limit reached when both neutralinos are moving in the same direction perpendicular to the beam direction. To passing the cut on $E_T^{\rm{miss}}$, a large mass splitting is required. Similar arguments hold for the momenta of two $b$'s. In the rest frame of $\tilde{b}_1$, the momentum of $b$ is around $\Delta m$. To passing the stringent cut with $p_T(b_1) > 130$~GeV, one basically requires to have $\Delta m \gtrsim 130$~GeV if the sbottoms are not moving too fast in the center of mass frame and there is no additional jets to boost sbottoms in the transverse direction. 

The story changes if there is an additional ISR jet with a large transverse momentum. The two-$\tilde{b}_1$ system turns to move in the transverse direction opposite to the ISR jet, as depicted in the right panel of Fig.~\ref{fig:isr}. For an ISR jet with a transverse momentum $p_T(j_1) < m_{\tilde{b}_1}$ and for a small mass splitting $\Delta m$, the sum of neutralino's transverse momenta is around $p_T(j_1)$ and hence can pass a stringent cut on the missing transverse momentum. For the two $b$'s in the final state, one can have the leading $b$-jet with a transverse momentum as large as $(1+ p_T(j_1)/m_{\tilde{b}_1})\Delta m$ if one sbottom does not move in the transverse direction, and the summation of the $p_T(b_1)$ and $p_T(b_2)$ to be $(2 + p_T(j_1)/m_{\tilde{b}_1})\Delta m$. At hadron colliders, the pair-produced sbottoms also have non-negligible velocities in the transverse direction. As a result, the $b$ jets produced from sbottom decays can easily pass the minimum kinematic requirement like $p_T(b_1) > 25$~GeV. However, it is very rare to pass the very stringent cut like $p_T(b_1) > 130$~GeV, imposed in the current ATLAS search~\cite{Aad:2011cw}, for the squeezed spectrum.  So, we propose a new search strategy to cover the squeezed mass spectra with a small value of $\Delta m$ by requiring the following three basic objects in the final state: 
\bit
\item \emph{One energetic non $b$-tagged jet from the initial state radiation.}
\item \emph{A large transverse missing energy.}
\item \emph{A $b$-tagged jet with a modest transverse energy.}
\eit
On top of those three basic requirements, we will also explore other kinematic variables to further improve the search limit of squeezed sbottom spectra.

Since the LHC is currently running with 8 TeV center-of-mass energy, all our simulation and estimation will be based on the 8 TeV LHC. We simulate signal and background events using \texttt{MadGraph5}~\cite{Alwall:2011uj}, and shower them in \texttt{PYTHIA}~\cite{Sjostrand:2006za}.  We use \texttt{PGS}~\cite{PGS} to perform the fast detector simulation, after modifying the code to implement the anti-$k_t$ jet-finding algorithm with the distance parameter $R=0.4$~\cite{Cacciari:2008gp}. To avoid the double counting issue, the background events are simulated using a parton shower plus matrix element matched method with the MLM scheme~\cite{Alwall:2007fs} implemented in \texttt{MadGraph5}. The signal production cross section is normalized to be the value calculated at NLO+NLL~\cite{Beenakker:2011fu}. We choose the branching ratio of $\tilde{b}_1\rightarrow b\tilde{\chi}$ to be 100\%. The main backgrounds for our analysis contain $t\bar t$ in semileptonic and dileptonic channels, single top production plus jets, $Wbbj$ and $Zbbj$ (we call $V+bb$ backgrounds later).  The background production cross section for $t\bar t$ is normalized to have a k-factor 1.7, calculated approximately at NNLO~\cite{Kidonakis:2010dk}. In our studies, the leptonic decays of the top quarks contain $\tau^\pm$ leptons. For other backgrounds, we use the k-factor 1.1 for the single top background~\cite{Schwienhorst:2010je}, $1.7$ for the $Zbbj$ background~\cite{Frederix:2011qg} and $2.0$ for the $Wbbj$ background~\cite{Cordero:2009kv}.

\begin{figure}[h!t]
\begin{center}
\includegraphics[width=0.5\textwidth]{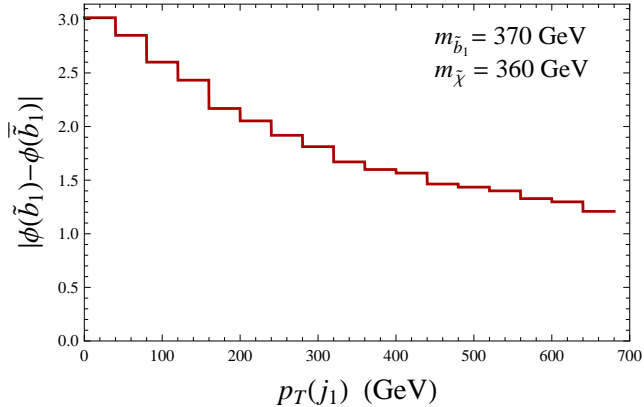}
\caption{The parton-level distribution of the averaged azimuthal angle difference of two sbottoms in terms of the ISR jet $p_T$.}
\label{fig:deltaphiparton}
\end{center}
\end{figure}

To illustrate the effects of the ISR jet on the kinematics of sbottoms, we show the averaged azimuthal angle difference, $|\phi(\tilde{b}_1) - \phi (\bar{\tilde{b}}_1)|$, as a function of the ISR jet $p_T$'s in Fig.~\ref{fig:deltaphiparton}. From this figure, one can see that for a small value of $p_T(j_1)$ the two sbottoms produced prefer to be back-to-back in the transverse direction and have the averaged azimuthal angle difference to be close to $\pi$, while for a large value of $p_T(j_1)$ they turn to move in the same direction with a smaller value of $|\phi(\tilde{b}_1) - \phi (\bar{\tilde{b}}_1) |$. The distribution of $|\phi(\tilde{b}_1) - \phi (\bar{\tilde{b}}_1)|$ confirms our schematic plots of the signal production mechanism in Fig~\ref{fig:isr}.
For the squeezed spectra, the missing transverse energy is approximately opposite to the ISR jet $p_T$ with the same magnitude. Therefore, the averaged $|\phi(\tilde{b}_1) - \phi (\bar{\tilde{b}}_1)|$ distribution as a function of the $E_T^{\rm miss}$ has a similar distribution as in Fig.~\ref{fig:deltaphiparton}. With a large missing energy cut, the two sbottoms are not back-to-back and hence the $m_{\rm CT}$ cut introduced for the case without ISR is not useful to cover the squeezed spectrum case.

\section{Additional Variables to Reduce the SM Backgrounds}
\label{sec:strategy}
Concentrating on the squeezed spectrum with a small mass splitting with $\Delta m = 10$~GeV, we require the following basic cuts to optimize the search
\bit
\item The leading non $b$-jet with $p_T(j_1) > 120$~GeV. 
\item A stringent cut on the transverse missing energy with $E_T^{\rm miss} > E_{T, \rm{min}}^{\rm miss}$, where $E_{T, \rm{min}}^{\rm miss}$ varies to optimize the search.
\item At least one $b$-jet with $p_T(b_1) > 25$~GeV and $|\eta| < 2.5$.
\item No leptons including $\tau^\pm$ with $p_T > 20$~GeV and $|\eta| < 2.5$.
\eit
Here, the last cut is introduced to cut the dominant $t\bar t$ background, which turns to have one or more charged leptons in the final state. To further reduce the backgrounds, we have found three more useful variables. 

\begin{figure}[h!t]
\begin{center}
\includegraphics[width=0.45\textwidth]{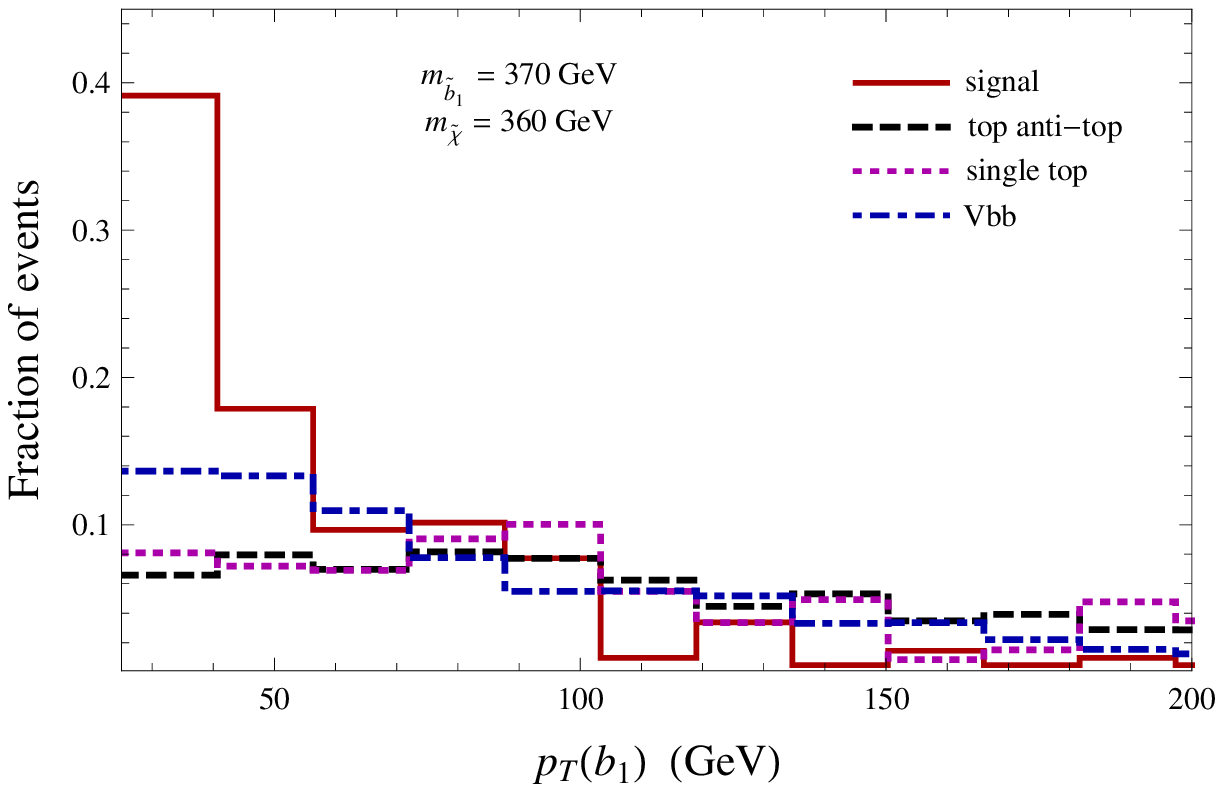} \hspace{3mm} 
\includegraphics[width=0.45\textwidth]{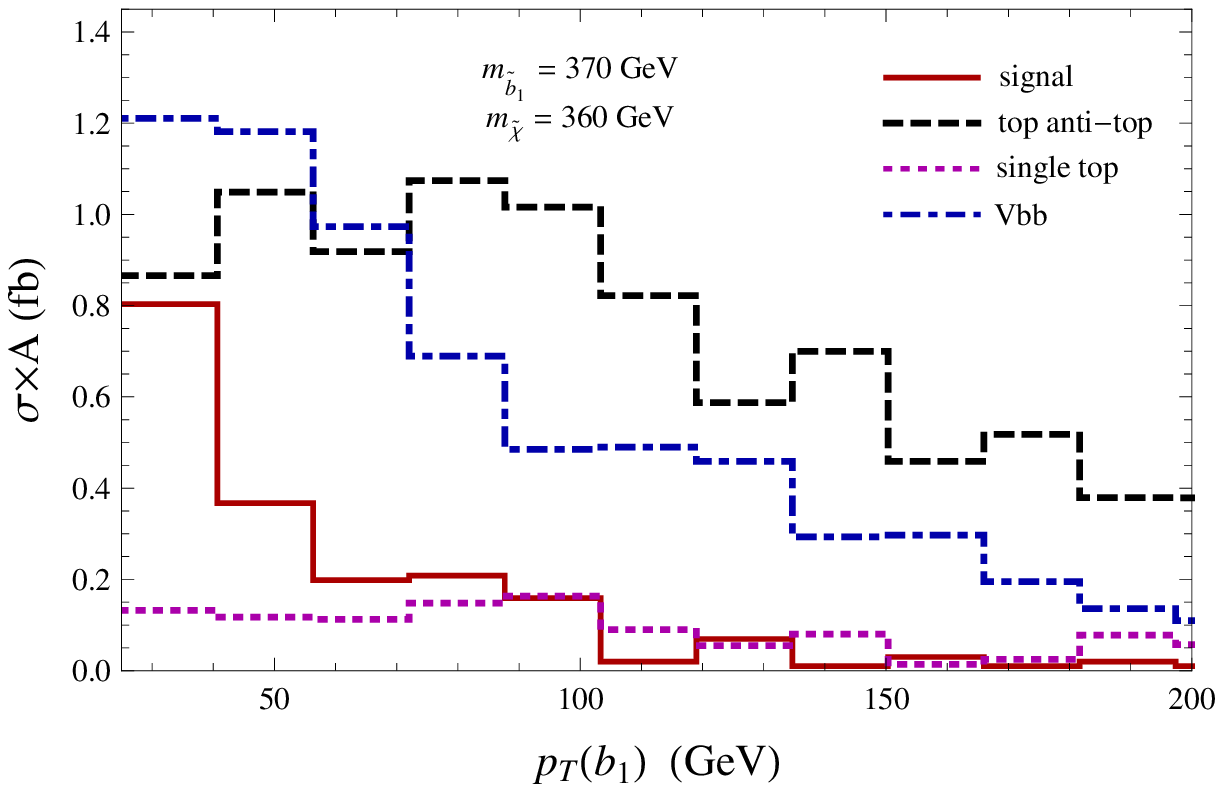}
\caption{Left panel: the normalized fractions of events for the signal and backgrounds in terms of $p_T(b_1)$. Here,  $m_{\tilde{b}_1} =370$~GeV and $m_{\tilde{\chi}} = 360$~GeV. A cut with $E_T^{\rm miss} > 400$~GeV has been imposed for all histograms. The signal acceptance is around 1.7\%. Right panel: the same as the left one but for the absolute cross section distributions after the $E_T^{\rm miss} > 400$~GeV cut.}
\label{fig:ptb1}
\end{center}
\end{figure}

The first variable is an upper limit cut on the leading $b$-jet. The reason for introducing this cut is because the leading $b$-jet $p_T$ from the signal is less energetic than that from backgrounds. From the top quark backgrounds, the $b$-jet comes from the top quark decay. With a sufficiently large missing transverse cut like $E_T^{\rm miss} > 400$~GeV in Fig.~\ref{fig:ptb1}, the top quarks are slightly boosted and have the decay product $b$-jet to be energetic. For the $W, Z + \mbox{hf}$ backgrounds, the leading $b$-jet prefers to be energetic to be summed together with the leading non $b$-jet to match the large $E_T^{\rm miss}$. For the signal, however, the missing transverse energy comes from two missing particles and both $b$-jet $p_T$'s are needed to compensate the difference between $E_T^{\rm miss}$ and the leading non $b$-jet $p_T(j_1)$. As shown in the left panel of Fig.~\ref{fig:ptb1} for the distributions of fraction of events, the sbottom signal does have almost all events with $p_T(b_1)$ below 110 GeV, while for backgrounds some fractions of events have $p_T(b_1)$ above 110 GeV. We will later vary the upper limit cut on $p_T(b_1)$ to improve our signal reaches. To have also a rough idea about the sizes of different backgrounds, we also show the absolute signal and background cross section distributions in terms of $p_T(b_1)$ in the right panel of Fig.~\ref{fig:ptb1}. 

\begin{figure}[h!t]
\begin{center}
\includegraphics[width=0.45\textwidth]{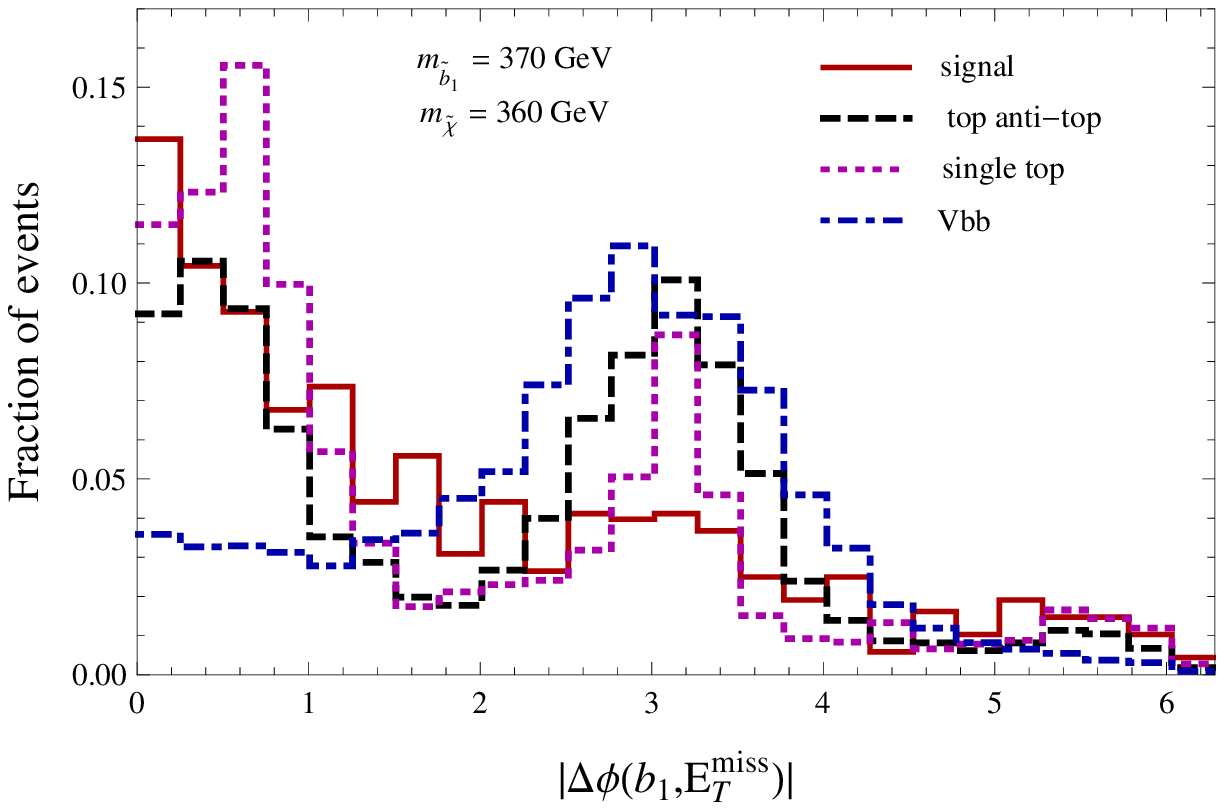}
\hspace{3mm}
\includegraphics[width=0.47\textwidth]{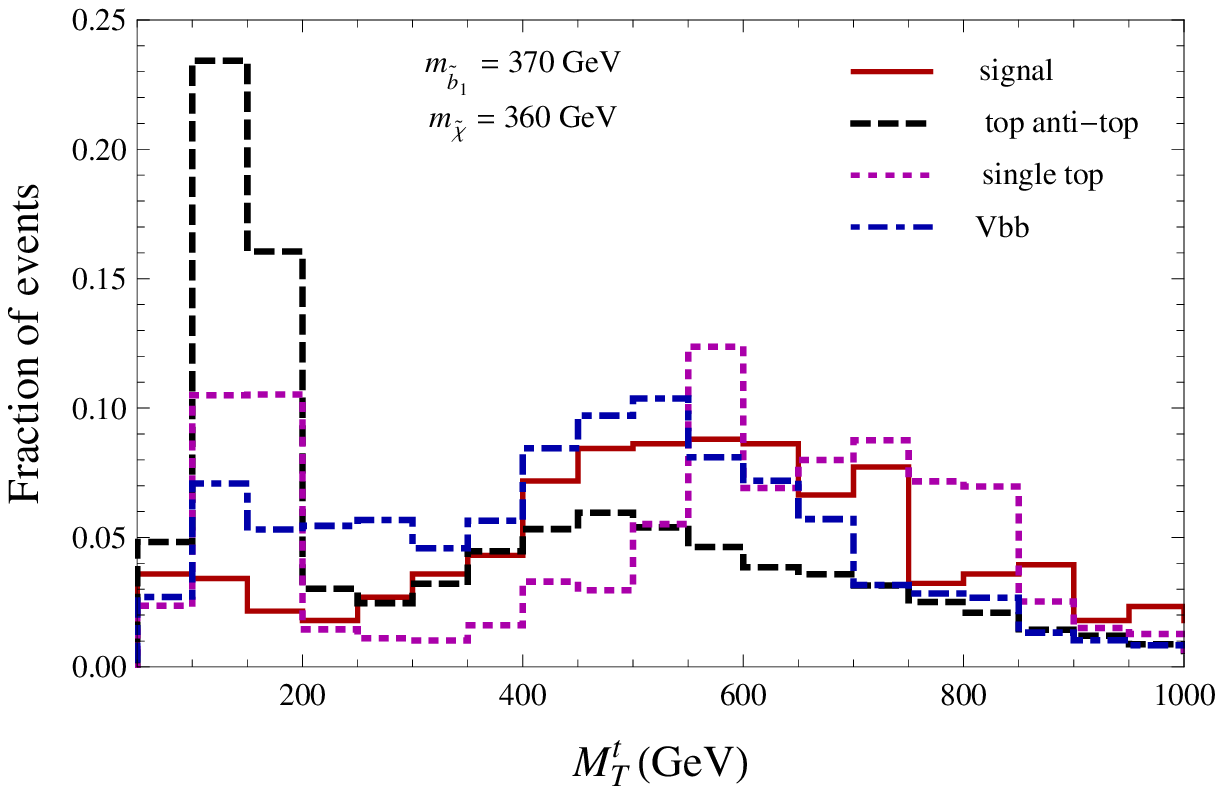}
\caption{Left panel: the $|\Delta\phi(p_T(b_1), E_T^{\rm miss})|$ distributions for the signal and backgrounds after imposing the cut $E_T^{\rm miss} > 250$~GeV. Right panel: the same as the left panel but for $M_T^t$ distributions.}
\label{fig:deltaphi}
\end{center}
\end{figure}

The second useful variable, $|\Delta\phi(p_T(b_1), E_T^{\rm miss})|$, is the azimuthal angle difference between the leading $b$-jet $p_T$ and $E_T^{\rm miss}$. For the $V+bb$ background, the leading $b$-jet prefers to have the same moving direction as the leading non $b$-jet, which is opposite to the transverse missing energy direction, so the $|\Delta\phi(p_T(b_1), E_T^{\rm miss})|$ distributions should peak at $\pi$ for this background. For the $t \bar t$ and single top backgrounds, the leading $b$-jet can align or anti-align with $E_T^{\rm miss}$, while for the sbottom signal the leading $b$-jet prefers to align with $E_T^{\rm miss}$. The reason has already been illustrated at the beginning of Section~\ref{sec:boost}. In order to have a large missing energy, the ISR jet will boost both sbottoms and make the bottom quarks move in the same direction as their corresponding neutralinos. We show the signal and background $|\Delta\phi(p_T(b_1), E_T^{\rm miss})|$ distributions in the left panel of Fig.~\ref{fig:deltaphi} after imposing a cut on the missing transverse energy with $E_T^{\rm miss} > 250$~GeV, the actual value of this cut does not change the general features of signal and background distributions, as far as $\Delta m$ is kept small.

The third variable is called $M_T^t$ or the top quark transverse mass. Since the $t \bar t$ background is one of the dominant backgrounds, reducing this background can definitely improve the final reach for sbottoms. It turns out that the dominant $t \bar t$ background is in the semi-leptonic channel with the lepton missing. So, this $M_T^t$ is defined specifically for this part of backgrounds.~\footnote{A similar but more complicated strategy for reducing the dileptonic $t \bar t$ background for the stop search, see Ref.~\cite{Bai:2012gs}.} We define this variable as
\beqa
M_T^{t} = \sqrt{ \left[ E_T(j_a) + \sqrt{ (E_T^{\rm miss})^2 + M_W^2 } \right]^2 - \left| {\bf{p}_T}(j_a) + {\bf{E}_T^{\rm miss}}\right|^2  } \,,
\label{eq:MTt}
\eeqa
where $j_a$ is one of the three leading jets with $p_T > 25$~GeV including the leading $b$-jet (or the leading two $b$-jets). We choose $j_a$ to be the one not belonging to the pair of jets with the smallest invariant mass among the three pairs. Since the lepton in the semi-leptonic $t \bar t$ background is missing to provide a large transverse missing energy with the neutrino, we use the $W$ gauge boson mass for the missing particle. Noticing that there are two $b$-quarks in the final state at parton level for this background, one of the two $b$-quarks should come from the same top quark as  the missing $W$. Because of the $b$-tagging efficiency from the detector simulations, there are more events with just one $b$-jet in the final state with the other $b$-jet either not tagged or too soft or too forward to pass the basic jet selection. Our way of selecting the right jet together with the missing $W$ to calculate $M_T^{t}$ is based on the fact that the two jets with the smallest invariant mass are more likely from the hadronic top and the remaining one is from the leptonic top. We show the $M_T^t$ distributions of signal and background events in the right panel of Fig.~\ref{fig:deltaphi}. As one can see from the right panel of Fig.~\ref{fig:deltaphi}, the majority of the $t\bar t$ background indeed has $M_T^{t}$ bounded by around the top quark mass, while the signal does have a good fraction of events with $M_T^t$ much above the top quark mass. So, a big fraction of the $t\bar t$ background will be reduced later by imposing a lower limit cut on $M_T^{t}$.

\section{Discovery Reach}
\label{sec:reach}

In this section we use the general search strategy with one $b$-jet, one hard ISR jet and a large $E_T^{\rm miss}$ plus the three variables in the previous section to study the sbottom discovery or exclusion reach. All our results reported here are for the 8 TeV LHC with an expected 20 fb$^{-1}$ luminosity.

We take two reference points in the $m_{\tilde{b}_1}$ and $m_{\tilde \chi}$ plane to study the optimization of the cuts.  To fully apply the search strategy developed in previous section for the squeezed spectrum, we take the reference point 1 to have $m_{\tilde{b}_1}=370$ GeV and $m_{\tilde \chi}=360$ GeV.  Since the optimized cuts for this reference point with $\Delta m =10$~GeV are not efficient to cover the parameter space with a larger $\Delta m$, we also consider a second reference point with $m_{\tilde b}=430$ GeV and $m_{\tilde \chi}=310$ GeV in order to cover the medium $\Delta m$ region between the squeezed spectrum and the large $\Delta m$ region.  We have verified that this point is outside of the reach of an extended analysis to $20$ fb$^{-1}$ using a search strategy similar as the one in Ref.~\cite{Aad:2011cw}.

To choose the best set of cuts for each of the reference points, we scan the parameter space for different cuts in $E_T^{\rm miss}$, $|\Delta\phi(b_1, E_T^{\rm miss})|$, $M_T^t$ and $p_T(b_1)$, and compute the numbers of signal ($S$) and background ($B$) events expected in 20 fb$^{-1}$ of luminosity at the 8 TeV LHC.  The details of the simulation have been depicted in Section~\ref{sec:boost}.  The Poisson probability that $B$ events purely from backgrounds would fluctuate up to at least $S+B$ events is given by {\tt ROOT}'s two-parameter $\Gamma$ function, which can be evaluated at non-integer parameters:
\beqa
p \ = \  \sum_{k=S+B}^\infty \frac{B^k}{k!} e^{-B} \ =\  {\tt TMath::Gamma}(S+B, B)~.
\eeqa
We translate this probability into a gaussian-equivalent significance ($\sigma$) in terms of standard deviations. This approaches $S/\sqrt{B}$ for a large signal and background. We select the best set of cuts for each reference point as the one that would maximize the significance and then apply the same cuts to other masses as well. 

\begin{table}[h!t]
\renewcommand{\arraystretch}{1.2}
\begin{center}
\begin{tabular}{|c|c|c|c||c|c|c||c|c||c|}
\hline
$E_{T}^{\rm miss}>$ & $M^t_T>$ & $|\Delta\phi(b_1,E^{\rm miss}_T)|$ & $p_T(b_1)<$ &  $\sigma_{t\bar t}$ & $\sigma_{tX}$& $\sigma_{Vb\bar b}$ &$\sigma_{B}$& $\sigma_{S} $& significance \\
(GeV)& (GeV) & $<$ & (GeV) & (fb) &(fb) &(fb)&(fb)&(fb)& (20 fb$^{-1}, 8$~TeV)\\
\hline
 430 & - & - & - & 8.2 & 0.3 & 5.9 &  14.4 & 1.5 &1.7\\
 430 & 200 & - & - & 5.0 & 0.3 & 5.5 &  10.8 & 1.4 & 1.9\\
 430 & 200 & 1.8 & - & 2.3 & 0.2 & 1.1 &  3.6 & 1.0 &2.4\\
 430 & 200 & 1.8 & 100 & 1.1 & 0.1 & 0.7 &  1.9 & 1.0 &2.9\\
\hline
\end{tabular}
\end{center}
\caption{Optimized cuts for the reference point 1 with $m_{{\tilde b}_1}= 370$~GeV and $m_{\tilde \chi}= 360 $~GeV.  The branching ratio of $\tilde{b}_1 \rightarrow b\tilde{\chi}$ is chosen to be 100\%. The last column represents the significance for each set of cuts expected at 20 fb$^{-1}$ of the 8 TeV LHC.}
\label{rp1}
\end{table}

To understand better the effects of different cuts on the signal and backgrounds, we present in Table~\ref{rp1} how the significance increases as the different cuts are imposed in succession for the reference point 1.  The basic cuts on other objects in the final state have been imposed and described at the beginning of Section~\ref{sec:strategy}. Besides vetoing any event containing at least one charged lepton (including tau's), we begin with an overall stringent cut of $E_T^{\rm miss} > 430$ GeV to boost the sbottom system. In agreement with the analyses in the previous section, a cut inspired in the right panel of Fig.~\ref{fig:deltaphi} of $M_T^t >  200$ GeV reduces the main $t\bar t$ background by about one half.  We then see that a cut in the angular variable with $|\Delta\phi(b_1,E^{miss}_T)|<1.8$ discards most of the $Vbb$ background and reduces one half of the $t\bar t$ and single-top backgrounds $tX$, as expected from the left panel in Fig.~\ref{fig:deltaphi}. A final upper-limit cut in the transverse momentum of the leading b-jet of $p_T(b_1)<100$ GeV further reduces the total background by around one half while not affecting the signal, in concordance with the analysis in Fig.~\ref{fig:ptb1}.  The combination of all these cuts enhances the significance for this reference point from $1.7$ to $2.9$ standard deviations. In our current study, we have neglected potential systematic errors from a realistic experimental search. However, since the three variables used in this table can increase $S/B$ from 0.1 to 0.5, we anticipate that our variables  are also useful to suppress the systematic errors for an experimental study.  

\begin{table}[h!t]
\renewcommand{\arraystretch}{1.3}
\begin{center}
\begin{tabular}{|c|c||c|c|c||c|c||c|}
\hline
$E_{T}^{miss}>$ & $M_{T}^t >$  &  $\sigma_{t\bar t}$ & $\sigma_{tX}$& $\sigma_{Vb\bar b}$ &$\sigma_{B}$& $\sigma_{S} $& significance \\
(GeV)& (GeV) &   (fb) &(fb) &(fb)&(fb)&(fb)&  (20 fb$^{-1}, 8$~TeV) \\
\hline
 270 & -  & 87.1 & 5.7 & 52.2 &  145.0 & 5.9 &2.2\\
 270 & 200 & 50.1 & 5.0 & 47.5 &  102.6 & 5.4 & 2.4\\
\hline
\end{tabular}
\end{center}
\caption{ The same as Table~\ref{rp1}, but for the reference point 2 with $m_{{\tilde b}_1}= 430$~GeV and $m_{\tilde \chi}= 310$~GeV. Due to a larger mass gap compared to the reference point 1 in Table~\ref{rp1}, cuts of $|\Delta\phi(b_1,E^{\rm miss}_T)|$ and $p_{T}(b_1)$ are found to be not useful for this point.}
\label{rp2}
\end{table}

For the reference point 2, where the mass gap is $\Delta m = 120$ GeV, the upper cut on $p_T(b_1)$ is not so useful because the signal events turn to have large values of $p_T(b_1)$. Similarly, a large missing transverse energy can be obtained without boosting sbottoms too much. The angular cut on $|\Delta\phi(b_1, E^{\rm miss}_T)|$ is not a characterization of the signal any more, since for this case the $b$ does not need to be aligned with $\tilde \chi$ in the lab system. Therefore, we don't use those two new variables: an upper cut on $p_T(b_1)$ and a lower cut on $|\Delta\phi(b_1, E^{\rm miss}_T)|$. Although the background events still have a peak structure for $2.5 \leq |\Delta\phi(b_1,E^{\rm miss}_T)| \leq 3.5$ as the distribution in the left panel of Fig.~\ref{fig:deltaphi}, we have found that imposing a cut with $2.5 \leq |\Delta\phi(b_1,E^{\rm miss}_T)| \leq 3.5$ can not increase the discovery sensitivity. For the transverse momentum of the leading $b$-jet, we have found that it does not have a clear upper bound that can differentiate the signal from backgrounds.  For this middle size of mass gaps, we have found that the only cuts that increase the significance are vetoing any charged leptons, a sufficiently large missing transverse momentum cut, and a lower cut on $M_T^t$ to reduce the top background.  We show in Table~\ref{rp2} the optimal cuts for the reference point 2 as well as the effects of the $M_T^t$ cut on reducing the $t\bar t$ backgrounds and increasing the final significance.

\begin{figure}[h!t]
\begin{center}
\includegraphics[width=0.45\textwidth]{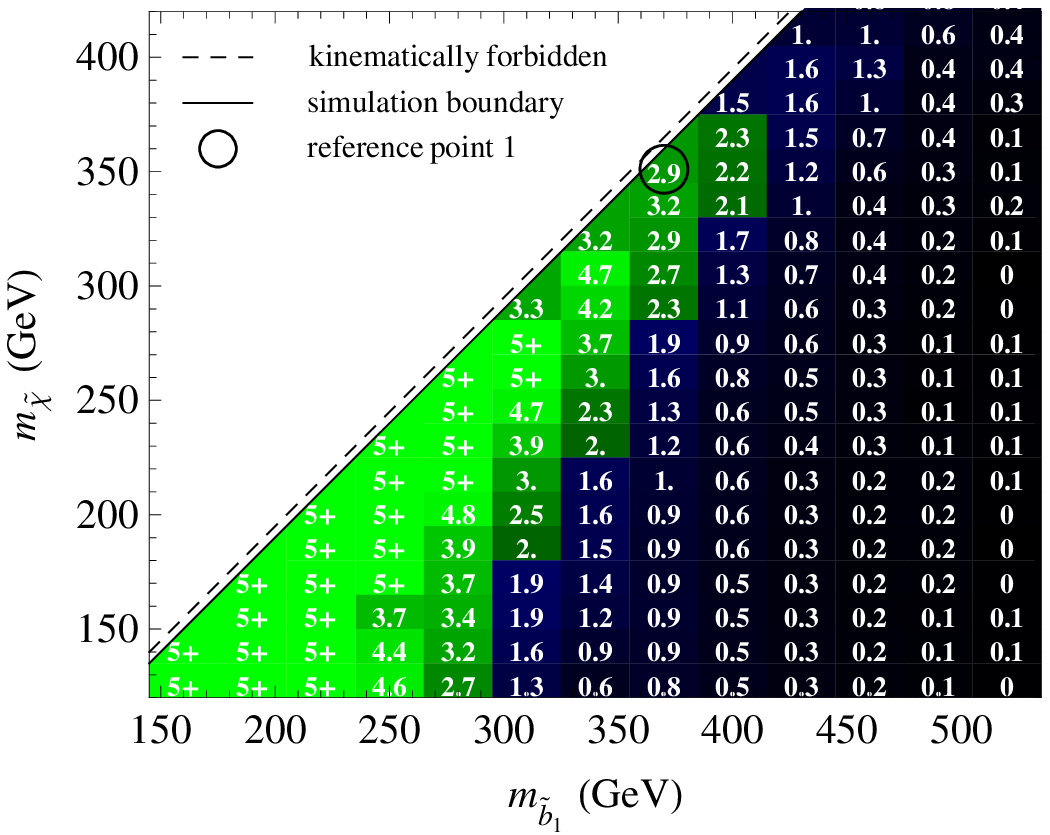}
\hspace{6mm}
\includegraphics[width=0.45\textwidth]{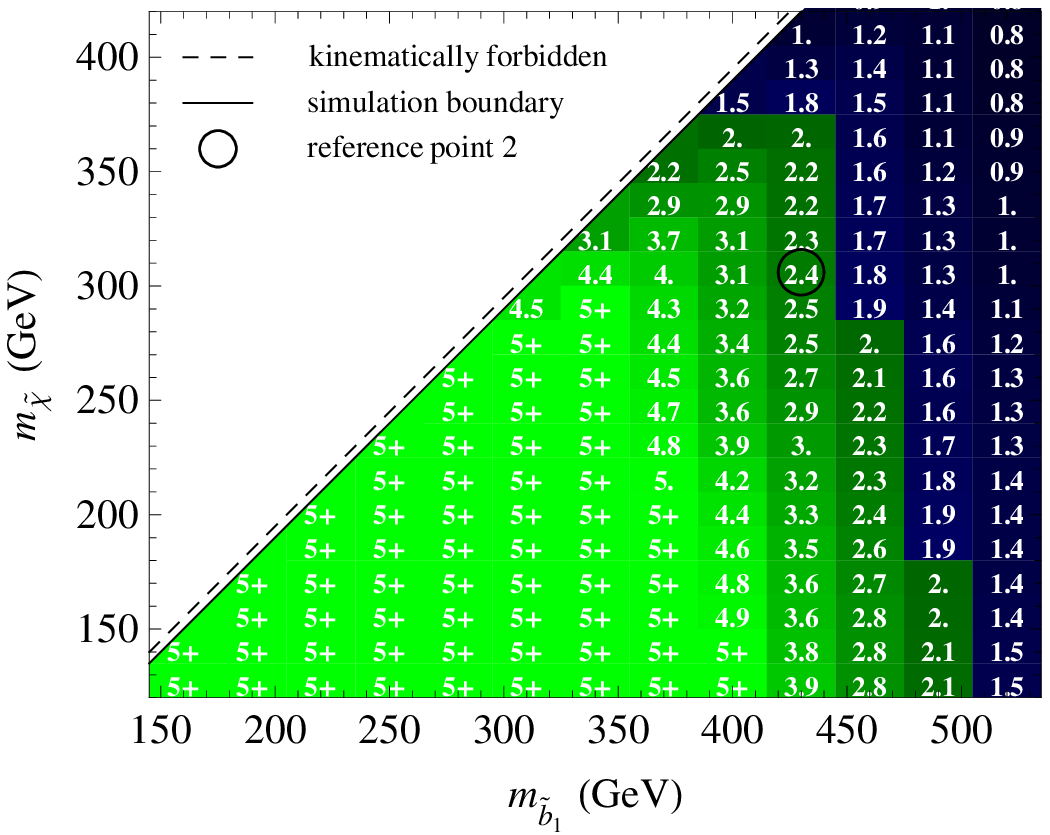}
\caption{Significance expected for 20 fb$^{-1}$ luminosity at the 8 TeV LHC for different sets of $m_{\tilde{b}_1}$ and $m_{\tilde \chi}$ using the optimal cuts found for the reference point 1 (left panel) and the reference point 2 (right panel). See Tables~\ref{rp1} and \ref{rp2} for detailed cuts.}
\label{results1}
\end{center}
\end{figure}

After finding the optimized cuts for these two reference points in Tables \ref{rp1} and \ref{rp2}, we apply them to other model parameter space of the $m_{\tilde{b}_1}$ and $m_{\tilde \chi}$ plane.  We show the discovery significance in the left and right panels of Fig.~\ref{results1} using the two reference cuts, respectively. In Fig.~\ref{results1}, the dashed line along the diagonal line is the kinematically forbidden limit, above which $\tilde{b}_1$ can no longer have a two-body decay into $\tilde{\chi}$ plus a $b$ quark at the parton level. Since the existing Monte Carlo programs do not cover this highly squeezed region precisely, we only simulate the parameter region with $\Delta m \ge 10$~GeV in our studies. The simulation boundary is shown as the solid line in the diagonal direction in Fig.~\ref{results1}. We leave the discussion of the extremely squeezed region with $5 \lesssim \Delta m < 10$~GeV in the next section.

From the left panel of Fig.~\ref{results1}, we see that the optimized search strategy for the reference point 1 with  $\Delta m = 10$ GeV produces an abrupt enhancement in a region which is parallel and close to the $\Delta m =10$ GeV line.  This can be understood from the set of cuts designed for this point: the upper bound on $p_T(b_1)$ and the cut on $|\Delta\phi(b_1,E^{\rm miss}_T)|$ differentiate signal from background only when $\Delta m$ is sufficiently small.  On the other hand, if $\Delta m$ is too small, a larger boost from the ISR jet is required to increase the missing transverse energy to reduce the backgrounds. The signal production cross section decreases as the ISR jet increases, so the significance also becomes worse. The actual reduction on significance depends on the sbottom masses. For a heavier ${\tilde b}_1$, a larger reduction of significance on the diagonal boundary of the left panel of Fig.~\ref{results1} is anticipated. As one can see from this panel, the highly squeezed region has a better coverage than the region with a large splitting. At 95\% C.L., one can exclude $m_{\tilde{b}_1}$ up to 400 GeV even when the mass splitting is as small as  $\Delta m \approx 10$~GeV.

The right panel in Fig.~\ref{results1}, on the other hand, has a slightly different pattern due to different reference cuts.  In this plot, for a fixed $m_{{\tilde b}_1}$ the significance decreases monotonically as $m_{\tilde \chi}$ increases. Because no cuts  on $p_T(b_1)$ and $|\Delta\phi(b_1, E^{\rm miss}_T)|$ have been applied for the reference point 2, we don't anticipate a similar pattern as the left panel. From the right panel, we can see that even though the set of cuts for the reference point 2 is not optimized for the highly squeezed region, a wide range of parameter space with a small $\Delta m$ can still be covered. Furthermore, this set of cuts can also cover the parameter region with a large $\Delta m$. Our results show that for a large $\Delta m$ above 300 GeV one can exclude $\tilde{b}_1$ up to 500 GeV at 95\% C.L. We also want to emphasize that the set of cuts for the reference 2 is not optimized for the large $\Delta m$ region. The search strategy used in the  ATLAS existing search~\cite{Aad:2011cw} by requiring two hard $b$-jets is the right one to cover this region.

Since both panels in Fig.~\ref{results1} have used the general search strategy by requiring 1 or 2 not-so-hard $b$-jets with one energetic ISR jet and a large $E_T^{\rm miss}$, it is not surprise that both can cover the highly squeezed spectra. Comparing the two results from the left panel and the right panel, we can see that the cuts used for the reference point 2 can cover most the squeezed spectra, while the cuts for the reference point 1 can have a better limit for the extreme cases with $\Delta m \approx 10$~GeV. 

\begin{figure}[h!t]
\begin{center}
\includegraphics[width=0.60\textwidth]{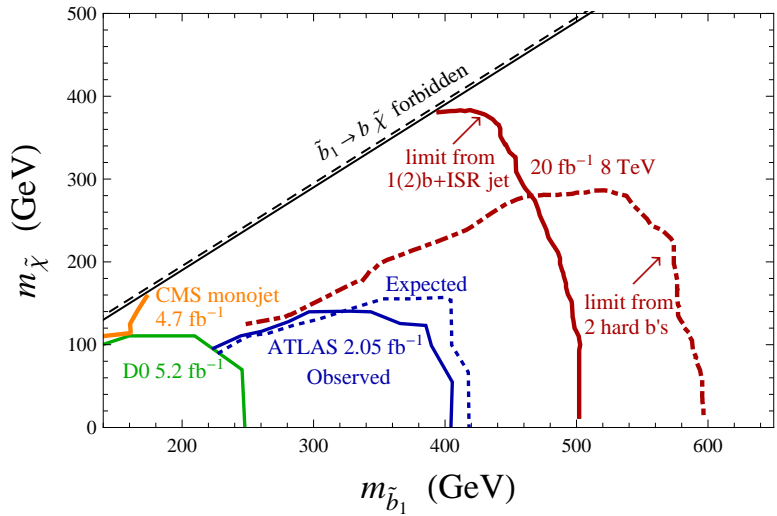}
\caption{The projected 95\% C.L. exclusion limits from our search strategy are shown in the red solid line for the 8 TeV LHC at 20 fb$^{-1}$. The existing search strategy by requiring two hard $b$-jets are shown in the red dot-dashed line with the same luminosity. All other limits from existing experimental searches are also at 95\% C.L.}
\label{results2}
\end{center}
\end{figure}

In order to compare our results to a search strategy like the one used in Ref.~\cite{Aad:2011cw}, we have also calculated the projected 95\% C.L. exclusion limit by cutting the first and second leading $b$-jet $p_T$'s to be above $170$ and $80$ GeV respectively, and requiring $E^{\rm miss}_T \geq 200$ GeV with all other cuts including the cut on $m_{\rm CT}$ left the same. The result is shown in the red dot-dashed line in Fig.~\ref{results2}. To combine the reach limits from using the two reference cuts in Fig.~\ref{results1} we take the larger significance for a given $m_{\tilde{b}_1}$ and $m_{\tilde\chi}$. The 95\% C.L. exclusion limit is shown in the red solid line in Fig.~\ref{results2}. Comparing the regions covered by our search strategy and the existing search strategy, we have found that a wide squeezed spectrum region, which is not covered by the existing strategy, is now covered by our proposal. Interestingly, the existing and our new search strategies are complimentary to each other in a sense that the existing search strategy has a better coverage for the large $\Delta m$ region. 

In Fig.~\ref{results2}, we also show existing experimental limits on the $m_{\tilde{b}_1}$ and $m_{\tilde \chi}$ plane. The blue solid and blue dotted lines are 95\% C.L. observed and expected exclusion limits from the ATLAS search results with 2.05 fb$^{-1}$ luminosity at the 7 TeV LHC~\cite{Aad:2011cw}. The green solid line is from the D0 search with 5.2 fb$^{-1}$ at the 1.96 TeV Tevatron~\cite{Abazov:2010wq}. In the orange solid line, we show the 95\% C.L. limit from the mono-jet search results from CMS with 4.7 fb$^{-1}$ luminosity at the 7 TeV LHC and using their $E_T^{\rm miss} > 350$~GeV cut~\cite{CMSmonojet}. The simple mono-jet search is not optimized to cover the squeezed sbottom-neutralino spectra.

\section{Discussion and Conclusions}
\label{sec:conclusion}
All our studies in this paper have the sbottom-neutralino mass splitting $\Delta m$ above 10 GeV. For the case of a smaller mass-splitting, the physics is more complicated but also interesting. When the mass splitting is reduced to be close but above the lightest $B$-hadron mass around 5.3 GeV, we anticipate that sbottom should decay into neutralino plus a single $B$-hadron. With the boost of the ISR jet, the signature of this region of parameter space should be one ISR jet, a large missing transverse momentum plus two single $B$-hadrons. The single $B$-hadron jet is different from the ordinary $b$-jet from QCD and can be tagged using a method beyond the standard $b$-jet tagging algorithm by requiring less tracks for the $b$-tagged jet. For the mass splitting below the mass of the lightest $B$-hadron, sbottom should have a multi-body decay into lighter hadrons plus neutralino. 

For the parameter region with a large $m_{\tilde{b}_1}$ and also a large $\Delta m$, one may think about studying the single sbottom production in association with a neutralino as $g\,b \rightarrow \tilde{b}_1\,\tilde{\chi}$. One may gain some production cross section by only producing one heavy particle in the final state. The final state is a mono-$b$ jet plus missing transverse energy. However, comparing to the pair-production of $\tilde{b}_1$, two additional reduction factors make the single sbottom production cross section very small. One is the small coupling among $b$-quark, $\tilde{b}_1$ and $\tilde{\chi}$ in comparison with the QCD coupling. The other one is the smallness of the fraction of the $b$-parton inside the proton. We have checked that the ratio of the single sbottom production cross section over the pair-production one is around $10^{-3}$ to $10^{-2}$ for $m_{\tilde{b}_1}$ from 600~GeV to 1~TeV, with a fixed the neutralino mass as 100 GeV. 

In summary, we have proposed a new search strategy to cover the squeezed spectrum with a small mass difference between sbottom and neutralino. The general search strategy of our proposal requires one hard ISR jet, one or two $b$-jets with a medium value of $p_T$'s and a large transverse missing energy. To further improve the search limit, we have found several useful variables including the azimuthal angle difference between the leading $b$-jet and $E^{\rm miss}_T$ and the top quark transverse mass $M_T^t$, defined in terms of a $b$-jet and the missing $W$ for the $t\bar t$ background. With a 20 fb$^{-1}$ luminosity at the 8 TeV LHC, we have found that our new search strategy can cover a wide range of parameter space with $\Delta m \lesssim 200$~GeV that would have not been covered by extending previous analysis. For a very small mass splitting with $\Delta m \approx 10$~GeV, we have found that the sbottom with a mass up to 400 GeV can be excluded at 95\% C.L.. Using the previous analysis strategy by requiring two hard $b$-jets, we estimate the reach for the sbottom mass to be around 600 GeV for $\Delta m \gtrsim 300$~GeV.

\subsection*{Acknowledgments}
We would like to thank Bart Clayton Butler,  Michael Peskin, Ariel Schwartzman, Daniel Silverstein and Peter Skands for useful discussion and comments. SLAC is operated by Stanford University for the US Department of Energy under contract DE-AC02-76SF00515. The author EA thanks Conicet for the special funding.

 \end{document}